# Orchestrating 5G Network Slices to Support Industrial Internet and to Shape Next-Generation Smart Factories

Tarik Taleb, Ibrahim Afolabi and Miloud Bagaa

*Abstract*—Industry 4.0 aims at shaking the current manufacturing landscape by leveraging the adoption of smart industrial equipment with increased connectivity, sensing, and actuation capabilities. By exploring access to real-time production information and advanced remote control features, servitization of manufacturing firms promises novel added value services for industrial operators and customers. On the other hand, industrial networks would face a transformation process in order to support the flexibility expected by the next-generation manufacturing processes and enable inter-factory cooperation. In this scenario, the 5G systems can play a key role in enabling Industry 4.0 by extending the network slicing paradigm to specifically support the requirements of industrial use cases over heterogeneous domains. We present a novel 5G-based network slicing framework which aims at accommodating the requirements of Industry 4.0. To interconnect different industrial sites up to the extreme edge, different slices of logical resources can be instantiated on-demand to provide the required end-to-end connectivity and processing features. We validate our proposed framework in three realistic use cases which enabled us highlight the envisioned benefits for industrial stakeholders.

## I. Introduction

Following the explosive spread of the Internet of Things (IoT) paradigm, industrial sites are equipped with an increasing number of connected devices, such as robots, sensors, and actuators [1]. This continuous evolution will yield a drastic transformation in the industrial landscape, allowing new production models and promising business opportunities. Indeed, there is an increasing impetus towards the development of dynamic and highly automated networked industrial environments, that are able to modify on-demand industrial production and support "product customization" with lower cost. On the other hand, new methodologies and mechanisms are strictly required to remotely verify the management process and check the ultimate quality of products. Next-generation industrial production systems should be able to seamlessly operate beyond the factory premises and provide new automation mechanisms by leveraging real-time information generated by industrial equipment.

However, the current industrial networks have been designed for static manufacturing processes where changes in the manufacturing workflow requires long maintenance operations. Furthermore, different traffic flows with potentially competing Quality of Service (QoS) parameters are forwarded over heterogeneous industrial environments. Last but not least, increased levels of security are required since current industrial actors are reluctant to share the production data with stakeholder due to the risk of losing sensitive data.

In this fervent area, we strongly believe that the advanced software-based network programmability features of the 5G mobile system [2] and other relevant technologies represent key enablers towards achieving success beyond Industry 4.0. By leveraging the adoption of Cloud Computing technologies, network can play a key role in next fourth industrial revolution, going beyond the conventional vision of bit pipes towards the management of dynamic traffic flows with enhanced innetwork processing. As a result, network slicing is gaining high momentum accounting for the increased capabilities to adapt network functionalities according to specific verticals.

In this paper, we present a platformized 5G system aiming at developing reconfigurable Industrial Internet environments for smart factories in order to support dynamic production processes, while optimizing incurred cost and enabling high level of remote monitoring and control. In this light, a holistic framework based on network slicing [3] is proposed for Industry 4.0. In this framework, the following key concepts are introduced:
(i) advanced network reconfigurability to support dynamicproduction processes;
(ii) seamless federation of edge cloud resources over multipletechnology and administrative domains;
(iii) enhanced radio communications to efficiently support wireless industrial-based requirements;
(iv) industrial equipment slicing so as to extend manageabilityup to the smart manufacturing devices and increase re-usability of production patterns; and
(v) control without ownership where the complexity of theprivate industrial infrastructure can be delegated to external operators. The proposed framework is validated in three different realistic use cases, which clearly point out the manifold advantages of the envisioned solution.

The paper is organized as follows. In Section II, an overview of current Industrial Internet solutions are described, especially highlighting the integration efforts towards IoT, Cloud Computing, and network softwarization. In Section III, we describe the proposed framework, pointing out the key driving concepts and the resulting architecture. Section IV includes potential use cases to validate the proposed approach, whereas Section V presents complementary industrial settings evaluation of the use case scenarios. In Section VI, conclusions and promising research challenges are drawn.

## II. Background on Industrial Internet

The success and sustenance of the fourth industrial revolution lies strongly in the adoption of emerging Information and Communications Technology (ICT) technologies, such as IoT, Cloud Computing, and network softwarization [4], which can be integrated into manifold manufacturing processes. In this section, we provide a brief overview of the current integration trends towards Industry 4.0.

The increased connectivity of smart devices, which is the main driver of the IoT paradigm, is rapidly extending its benefits across industrial factory environments. This enables the opportunity to establish cyber representations of physical industrial equipment through appropriate data digitalization, ensuring an advanced vision of the current production environments [5]. The actual enhancement in monitoring, control, diagnosis, and maintenance processes depends on the ability to effectively collect, analyze, and extract crucial information from multiple heterogeneous sources, potentially deployed over geo-distributed production plants. Therefore, the current industrial ICT infrastructures should gradually evolve to efficiently support this new trend, which represents a gamechanging paradigm for the competitiveness of industries over the next coming years.

Great efforts have been carried out to efficiently integrate Cloud Computing technologies within industrial domains and to enhance information processing. Cloud manufacturing [6] is a new multidisciplinary paradigm where configurable manufacturing resources are managed in a centralized way and provided as cloud services. In [7], a cloud platform represents the backbone of an evolved manufacturing system which is able to support flexible production. The increased capabilities offered by cloud technologies are exploited to translate the customers' requests into specific relevant production tasks. Another application scenario deals with cloud robotics [8], whereby the highly complex tasks involved in computing processes are moved from physical devices to the cloud platform. This approach notably increases the potential runtime operations of robots while providing on-demand reconfigurations capabilities. In this vein, accounting for the strict requirements in terms of remote control and bearing in mind that remote cloud data centers can suffer from high latency has pushed for the adoption of edge computing [9]. However, the distribution of computing over heterogeneous data centers for industrial applications is still at its infancy.

On the other hand, to provide the desired interworking capabilities of distributed industrial systems, several protocols have been proposed. In particular, beyond conventional shortrange communication protocols for industrial wireless sensor networks, Low Power Wide Area (LPWA) networks, such as Lora and Sigfox, can provide long-range communications with low energy consumptions [10]. Furthermore, within the efforts to standardize the radio interface for next-generation 5G systems, novel cellular-based technologies have also been investigated to support massive IoT applications, such as Narrowband-IoT [11]. However, supporting the latency and reliability requirements of industrial wireless applications still represents a major challenge and the coexistence of different radio technologies requires further analysis.

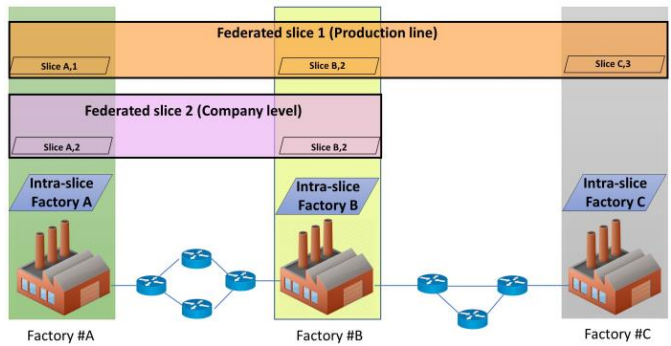

Fig. 1. Federations of industrial network slices over heterogeneous factory infrastructures.

In the context of industrial networks, Software Defined Networking (SDN) has also recently received notable attention accounting for the increased flexibility and programmability of networks. By decoupling data and control traffic, a logically centralized controller can on-demand modify traffic flows according to industrial application requirements. In [12], SDNbased industrial networks are proposed to support dynamic production processes. In particular, by enabling runtime optimization and disabling redundant production lines, remarkable energy savings can be achieved with respect to traditional industrial networks.

Network Function Virtualization (NFV) is the technology that allows the virtual representation and deployment of normal physical network functions as Virtual Network Functions (VNFs). NFV, when combined with SDN and Cloud Computing, can provide a strong impetus for the rise of network slicing solutions in industrial networks. However, the current proposed solutions for 5G do not specifically take into account the peculiarities of industrial networks, such as collaboration over heterogeneous factory environments, industrial equipment sharing, and control without ownership. This work presents an innovative framework where the network slicing paradigm has been extended to comprehensively support a broad range of industrial use cases.

## III. Network Slicing Orchestration for Industrial Internet Environments

Our framework aims at supporting the creation and lifecycle management of multiple network slices over factory infrastructures in order to enable unprecedented level of flexibility, scalability, and data manageability for industrial processes. Indeed, next-generation factories would be able to support extremely different traffic flows with potentially competing requirements in terms of performance, reliability, and security. Whilst a single monolithic architecture fails to support these novel industrial scenarios, network slicing paradigm in the 5G landscape can represent the key enabler for Industry 4.0 to finally boost the desired interworking among industrial processes. In Fig. 1, we present our vision where appropriate network slices can be created on-demand over factory infrastructures. Each slice provides the desired logical connectivity and contains both VNFs and Virtual Application Functions (VAFs) to ensure a desired data processing capability. Since production processes are usually distributed over multiple production plants, specific federation processes are required to efficiently enable cooperation over factory boundaries.

*A. Key Concepts*

In this subsection, we describe the main key concepts behind the proposed framework to enable an effective end-to-end industrial interworking.

*a) Reconfigurable Network Slices for Dynamic ReProgrammability of Industrial Production Workflows:* A main feature of next-generation factories will be the capability to dynamically reconfigure and reprogram its production processes and relevant control data flows. This represents a drastic shift with respect to the current industrial environments, where critical control/data streams are created and updated rather infrequently. Since current industrial networking solutions do not take into account this aspect, advanced network softwarization technologies allow coping with the increasing flexibility of industrial production systems. Network slices, used for the manufacturing of a product, can be configured on demand or based on predictions of the changing specific requirements of industrial flows, aiming at constantly providing the desired QoS for robotic machines, sensors, actuators, and control loop processes. Furthermore, the management of these network slices can ensure the desired isolation between data flows belonging to potentially competing companies (i.e., consumers of these slices).

*b) Federation of Heterogeneous Edge Slices over Cross-border Factory Environments:* The strict time requirements within industrial production processes require the abilities to host application functionalities and therefore extend the instantiation of slices up to the extreme edge of the network. This approach allows the provisioning of data processing capabilities closer to the manufacturing environments and ensure fast control loops. On the other hand, the key challenge is represented by the heterogeneity of edge data centers, from both an administrative and technological point of views. Our framework is designed to interact with defined private/public edge cloud infrastructures, to establish appropriate slice federations. A peculiar feature of the proposed system, when compared to classic cloud federations, concerns the possibility to update the slice of edge resources within short time intervals, so as to meet the changing conditions of dynamic production processes.

*c) Advanced Radio connectivity for Industrial Equipment:* Wireless interconnectivity of industrial equipment is gaining high momentum within industrial environments. Indeed, mobile wireless devices can notably enhance the flexibility of manufacturing processes compared to the traditional automation approach implemented with wired components. On the other hand, the wireless communication requirements, in terms of energy efficiency, latency and reliability, can vary significantly ranging from low power and long range communications to ultra-reliable low latency communications needed for real-time control. The first set of requirements is best met with Narrow-Band IoT (NB-IoT) while the latter requires 5G Ultra-Reliable Low-Latency Communications (URLLC). Hence, the Radio Access Network (RAN) for Industrial Internet is likely to have multiple Radio Access Technologies (RAT). RAN can support multiple core network slices and relevant RAN resources can be also sliced in a dynamic manner for multiple services. RAN virtualization is significantly different from the virtualization of core networks and introduce remarkable challenges in the allocation of limited spectrum resources. However, still notable efforts are required to design optimization techniques able to dynamically allocate radio resources among RAN slices [13]. Our framework aims to leverage and support the coexistence of multiple wireless technologies to fulfill the requirements of competing industrial use cases.

*d) Industrial Equipment Slicing:* To effectively support the reprogrammability of industrial processes, the network slices can be extended to involve the industrial equipment and their relevant controllers. This requires the adoption of virtualization technologies to dynamically update the control behavior and to enable the efficient sharing among multiple tenants. Appropriate sandboxing strategies are indeed required to provide application suite and customizable execution environments . In this vein, some preliminary efforts are proposed in [14], where Programmable Logic Controllers (PLCs) are virtualized in industrial control system by leveraging real-time hypervisor. The resulting prototype shows the feasibility to decouple the logic and control capabilities from the I/O components. Indeed, the support of deterministic requirements are at its infancy in network softwarization and considerable margin of improvements can be achieved through convergent computing and networking approaches. Nevertheless, industrial equipment slicing offers promising advantages since the same industrial device may be associated to multiple virtualized controllers which can steer their traffic flow through specific slices. In this way, different security and privacy contexts can be ensured according to the involved industrial processes, as well as the opportunity to update at run-time some controlling features while guaranteeing the overall service continuity.

  *e) Controlling without ownership:* Highly virtualized industrial environments requires appropriate skills in network softwarization, which can go beyond the expertise of current industrial operators and can represent a barrier towards the adoption of the envisioned smart factory models. To overcome this issues, the concept of controlling without ownership can be enabled to move the burden of complex network management away from the industrial owners to expert third-party operators. Indeed, by increasing the abstraction of physical infrastructure, external administrators can efficiently manage the configuration of proprietary industrial equipment by lever-aging standardized interfaces. Furthermore, this strategy can stimulate the creation of micro mobile operators able to tailor end-to-end network management for specific industrial environments and services [13]. In an industrial environment, besides external core networks, there can be also a local core network and edge computing servers for ultra-low latency applications. A local core may be needed to keep some of the data completely private - not leaving the industrial site at all.

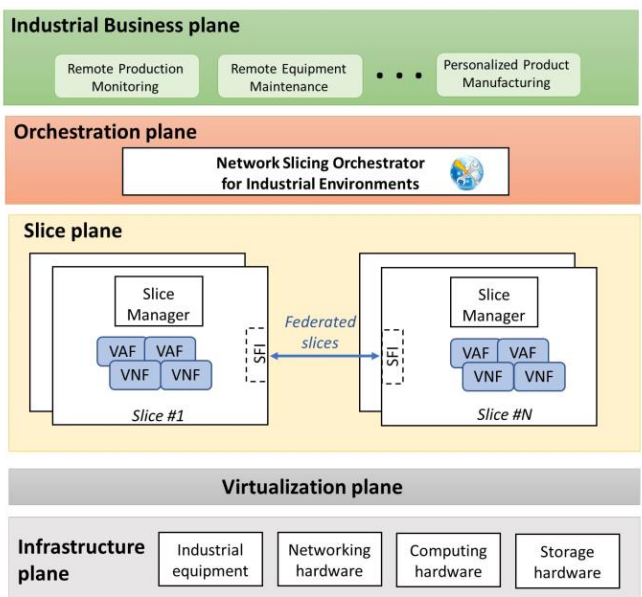

Fig. 2. Framework architecture for the orchestration of industrial network slices.

Furthermore, accounting for the deployment of a large number of software services across multiple cloud and edge data centers, the complexity of the management system can notably increase. Therefore, a centralized orchestration system may not be able to coordinate the virtual industrial functions and take the appropriate decisions in due time, as it needs to handle a large amount of runtime operations. Some specific management processes can then be delegated within the specific slice and relevant controllers can be hosted as virtual function (VF) instances within cloud edge infrastructure. In this way, sensitive manufacturing data can also be monitored and analyzed within the factory boundaries and can be exchanged with external stakeholders only after being appropriately processed.

*B. Architecture*

In this subsection, we present the high-layer architecture (Fig. 2) of the proposed framework which is intended to efficiently orchestrate network slices over multiple industrial environments. The *Infrastructure Plane* provides all the physical resources which are required to execute virtualized industrial processes. We point out that the involved resources go beyond traditional data centers, including not only physical computing, storage, and network components, but also covers the industrial equipment with sensing and actuation capabilities. The resources can be within the same administrative control or entirely belong to different administrators, where a combination of public/private domains is envisioned.

The *Virtualization Plane* refers to the tools and technologies required to provide a virtualization environment for hosting VNF and VAF instances. To enable this layer, hypervisor- and container-based virtualization technologies can be exploited to provide the execution environments for different software functions with varying requirements in terms of security and real-time constraints. Furthermore, the inclusion of industrial equipment in the lower layer would require novel abstraction schemes to be integrated into the virtualization plane. These schemes would allow the virtualization plane to also consider both the sensing and actuation capabilities of the industrial equipment, which can be offered as services with different levels of granularity.

The *Slicing Plane* refers to the industrial slices deployed to accommodate specific industrial use cases. Our framework aims to support end-to-end process, by enhancing industrial flows with in-network processing capabilities. Indeed, by leveraging edge cloud resources, not only VNFs, but also VAFs can be deployed closer to the industrial equipment, thus providing real-time control loop and traffic analysis. Since the operations can be extremely different within industrial slices, specific *Slice Managers* are instantiated to coordinate the lifecycles and interworking among slices' components. Further management services, such as reliability, security, and performance, can be supported according to the business requirements.

The slice federation over industrial domains can be ondemand enabled by leveraging specific interfaces, referred to as *Slice Federation Interface* (SFI). These interfaces can be appropriately configured to define the interworking among different slices, specifying the inbound/outbound traffic flows and the required data processing. These interfaces are of utmost importance in industrial domains, where the granularity and encryption of exchanged data is fundamental to ensure the protection of sensitive manufacturing information.

The *Orchestration Plane* represents the core of the proposed systems enabling on-demand slices over heterogeneous industrial domains. Industrial slices can be generated for multiple industrial use cases, such as monitoring, maintenance, and production. Furthermore, these slices can be extended up to the extreme edge , including industrial equipment when appropriate virtualization technologies are provided within relevant industrial domains. The orchestration system is in charge of guaranteeing different QoS constraints, according to relevant industrial traffic patterns, and on enabling network reconfigurability to support dynamic changes in industrial processes. To this end, the orchestrator interacts with specificdomain management components, such as SDN controllers, ETSI-based NFV management and orchestration (MANO) modules [15], and industrial controllers. For instance, by interacting with NFV MANO modules in charge of specific domains, the orchestrator can acquire and allocate predefined virtual resources to specific slices and enforce appropriate interworking between the cross-border slice virtualization.
The instantiation of federated slices can be an extremely challenging task and adequately resolving potential conflicts in resource allocation which can emerge among competing slices is an essential functionality the orchestrator must handle. To resolve this important functional challenge, the orchestrator can define appropriate admission control policies to dynamically evaluate new requests, present the available resources and enable interoperability of the newly instantiated slice(s) with the existing ones.

The *Industrial Business Plane* includes the tools and interfaces to allow industrial stakeholders to provide their requirements in the network slicing creation. For example, big companies can integrate their Enterprise Resource Planning (ERP) systems with the envisioned orchestration system to enhance the product manufacturing over industrial environments. The extended programmability of the orchestration system can indeed enable a broad range of use cases, as discussed in the next Section.

## IV. Industrial Use cases

Our vision of next-generation factories can represent a breakthrough in the current industrial context. First, the outsourcing of industrial production can introduce new business models, while guaranteeing the same product quality levels. Furthermore, the dynamic re-programmability of industrial processes can allow the customization of product manufacturing processes according to the requirements of producers/designers at a minimal cost. To clearly describe the advantages of the proposed framework, we present three exemplary use cases, which highlight how enhanced reconfigurability features of next-generation Industrial Internet can support novel industrial business scenarios.

### A. Remote Industrial Production Monitoring

Multi-national companies, such as electronic and automotive industries, have increasingly outsourced several stages of the overall production to reach economies of scale goals. The ultimate end-user product is typically a result of the assembly of manifold components produced by different third-party companies. However, to effectively ensure the overall product quality, each single component has to guarantee the expected quality levels. Therefore, new methodologies to monitor the production chain are highly required to keep one step ahead of potential issues in product quality, which can cause tremendous loss of revenues and significantly damage company's reputation. In this vein, it is significant to report the case of Samsung Galaxy Note 7: Samsung Electronics outsourced the production of the batteries of its Galaxy Note 7 to one of its subsidiaries. The batteries presented manufacturing defects, causing the explosion of the devices. The recalling of the products from Samsung generated a loss of billions of USD, not to mention the relevant impact on the reputation of the company.

To better explain the use case, we may consider the following scenario. An Industrial Product Company (IPC), called $IPC_A$, decides to outsource the production of some components of its product to a Contract Manufacturer ($CM_1$).

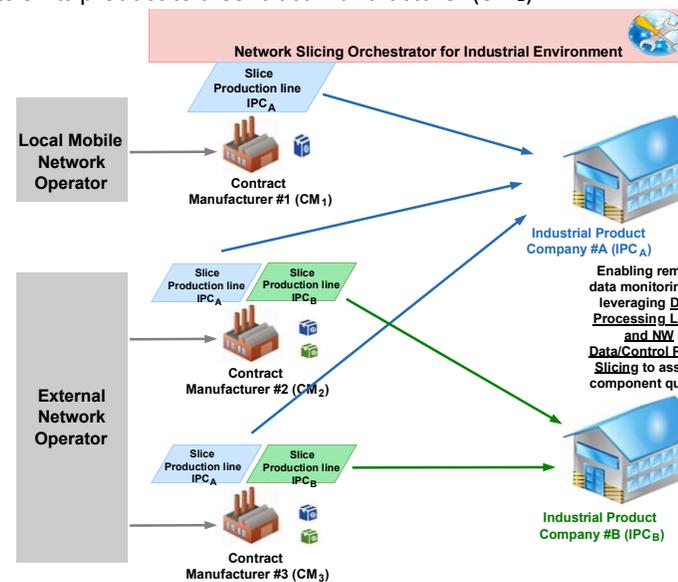

Fig. 3. Use case 1: Slicing for remote monitoring.

To reduce the risks associated with lack of product quality control, $IPC_A$ requires from $CM_1$ the capabilities to remotely monitor the outsourced production. In this vein, the proposed framework enables such "remote product quality control", by leveraging and extending the principles of network slicing and smart data management. Using the orchestration system described in Sec. III, $IPC_A$ will be offered an optimal slice equipped with the necessary VNFs and VAFs to remotely monitor the production line of its outsourced components. Depending on the agreement between $CM_1$ and $IPC_A$, the slice may offer the possibility of intervening with the production process when necessary. Only the information relevant to the production of $IPC_A$'s components must be shared with $IPC_A$, while ensuring the confidentiality of other sensitive data relevant to other customers of $CM_1$ or owned by $CM_1$. In Fig. 3, we capture this scenario, but in this case involving multiple CMs and IPCs. Fig. 3 presents the case whereby, the same IPC can outsource the production of multiple components to different CMs (e.g., $IPC_A$ subcontracting $CM_1$, $CM_2$, and $CM_3$)

Our framework is aimed at enabling smart access to data generated by industrial equipment during operational activities. By jointly leveraging and extending mechanisms and algorithms of network slicing and data processing at the edge, selected information provided by smart equipment can be made remotely available to their relevant vendors. In Fig. 4, an exemplary scenario is illustrated. The network equipment sold by a vendor $IEV_A$ can be used by different industrial manufacturer and exploited in manifold productions. The envisioned framework allows $IEV_A$ to deploy distributed slices, which contain all the networking and computing modules to have access to the necessary equipment data. Through and a CM can sign contracts with multiple IPCs (e.g., $CM_2$ being subcontracted by $IPC_A$ and $IPC_B$) simultaneously. Hence, the need for secure and tight isolation of traffic and data relevant to the productions of different IPCs is extremely crucial and can be provided using dedicated slices.

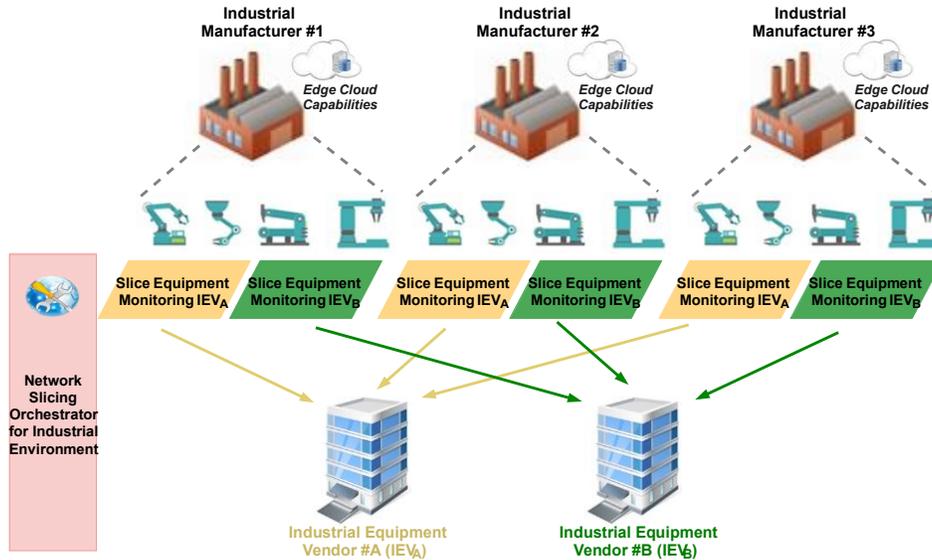

### B. Remote Equipment Maintenance

Industrial Equipment Vendors (IEVs) are demonstrating an increasing interest in monitoring their shipped equipment during the whole lifecycle. The data gathered about the operational behaviour of equipment can provide useful hints to identify possible shortcomings in equipment design and lead to several optimizations. Furthermore, the opportunities to interact with equipment even after sales can enable new services with additional revenues, such as remote maintenance and reconfigurability. However, many industrial companies are reluctant to make data generated by the equipment accessible to their respective IEVs. This comes as a result of the fear of a potential leakage of sensitive information related to their own products and production processes.

Fig. 4. Use case 2: Slicing for network equipment maintenance.

appropriate federations of the involved slices, the $IEV_A$ can uniformly gather and merge data originating from different industrial environments, and provide advanced maintenance services.

### C. Dynamic Industrial Manufacturing

Software-based programmability of next-generation factory infrastructure, in conjunction with smart general-purpose robots and industrial assets, can provide re-programmable industrial environments which have the potential to create novel manufacturing models in product development. Similar to cloud environment whereby computation and storage capabilities are offered on a pay-per-use basis, the same industrial infrastructure can be partially or totally rented to multiple external companies. In this way, the industrial infrastructure owner could enable on-demand manufacturing of any external company's products, according to a Smart-Factory-as-aService (SFaaS) model. This model can also promote the diffusion of small-medium IPCs, which can develop products without effectively owning the potentially expensive industrial assets. In such a case, the initial infrastructure investment will be drastically reduced for the IPCs, therefore allowing them focus on the product design by developing appropriate product manufacturing chain blue-prints.

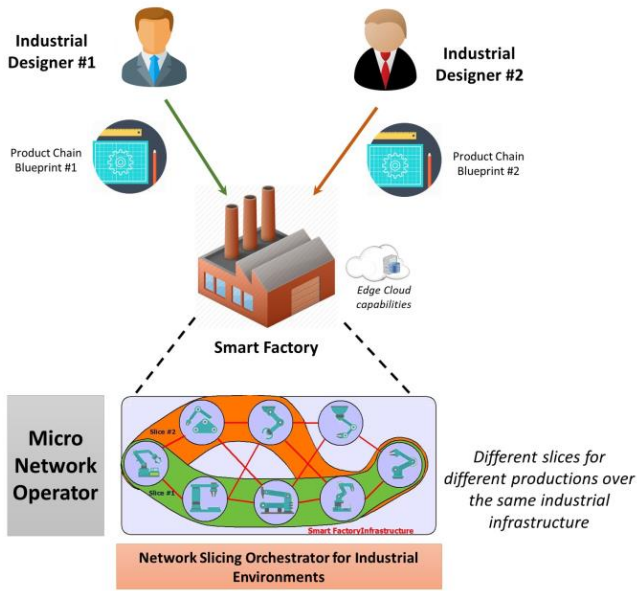

Fig. 5. Use case 3: Slicing for smart manufacturing.

In Fig. 5, we sketch a scenario where an Industrial Designer can provide a Product Chain Blueprint as inputs for the systems. Accounting for the envisioned requirements for product manufacturing, specific slices will be deployed to instantiate the relevant control modules and ensure networking among the required industrial equipment. Indeed, several robots can be interconnected to support efficient manufacturing along the envisioned production workflows and reconfigurable network slices can be created and updated to enable the relevant traffic flows with the desired QoS.

Product Owners (POs) to represent: (1) the IPCs as in the use case described in Section IV-A, (2) the IEVs as in the use case described in Section IV-B and in a way, (3) the industrial designer as in the use case described in Section IV-C. The second terminology is the Product Manufacturer (PM), which represents the (1) CM, (2) Industrial Manufacturer and (3) Smart Factory as in the three use cases, respectively.

A generic setup is possible for the three use cases since they all have something in common, which is the creation and use of network slices to connect a remote location to a smart factory site. The slices are created in order to either maintain an equipment, monitor a production process or dynamically manufacture products using different blueprints. Since the use of slices is common to all three use cases and each network slice would be utilized by the different POs to connect to the smart factory owned by a PM, it is vital to investigate the impact that already running network slices would have on the time taken to create new slices. The time taken by the system to instantiate a new set of connected VFs to form a new slice is referred to as the system's *Response Time* as shown in Figs. 6(a) and 6(b).

To capture the above setup scenario, we used two computers and a set of micro-computers (e.g., raspberry pis). The first computer is used to emulate the POs, with container instances running on it. Each container instance represents a PO willing to request the creation of a network slice for carrying out any of the use cases described in Section IV. The second computer represents the PM's server. It receives slice creation requests from POs, creates the slices and connects each PO with its respective slice. Each slice is composed of connected VFs running on the raspberry pis. The VFs are running on

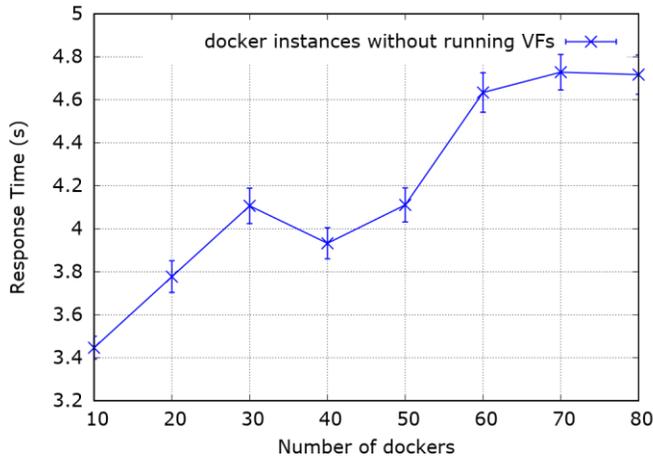

(a) Case of newly-created empty docker instances.

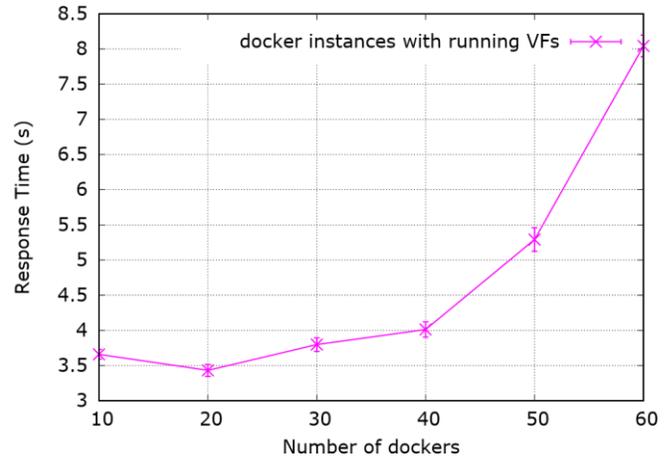

(b) Case of docker instances newly-created with actually running VNFs.

Fig. 6. System responsiveness in setting up additional new slices based on the number of already running slices.

## V. Evaluation Setup and Analysis

In this section, we present the results of the evaluation of our generic system setup. In our system setup, we considered the three use cases described in Section IV. In order to make the setup more generic and capture all the use cases, we introduced a set of terminologies. The first terminology is the docker instances deployed on the raspberry pis. In our setup, the raspberry pis are used to mimic lightweight micro-computers that are for instance mounted at every production line in a large-scale smart factory assembly plant. These lightweight micro-computers could be used: (1) to monitor the physical properties (e.g., temperature, sensitivity, and strength) of different product parts, which are currently on each line, (2) to capture data for maintenance purposes from each sophisticated manufacturing machine positioned at each line, and (3) to dynamically exchange information about the blueprints of any products whose production is about to begin.

For the aforementioned scenarios, we tested the responsiveness of the system in setting up new slices, i.e., how fast the system can instantiate a new set of connected VFs to form a new slice based on the number of already running VFs on the raspberry pis. The results are in two parts as shown in Fig. 6(a) and 6(b), respectively. In our setup, we used a total of 9 raspberry pis instantiating up to 80 docker instances on each, at every round in the experiment (i.e. 10 at a time, and a total of 10 rounds of experiments). Fig. 6(a) shows an overall progression of the system responsiveness in creating new empty docker instances to form a new slice with only computation and storage resources but no specific VFs running (i.e., deemed to be installed later by the respective product owner). In the figure, we notice that there are situations when the system's response times do not change much when increasing the number of running instances (e.g., 30 to 50 instances running). Whilst it is difficult to find a clear rationale beneath this performance, our investigation revealed that this could be as a result of the current state of the system at that particular point in time with respect to how less busy the processors are and other running jobs. On the other hand, in Fig. 6(b) that shows the impact of newly created docker instances with running VFs on the system's responsiveness, the trend is pretty clear and the impact of the running VFs are distinctive on the response times. Correspondingly, there is a noticeable increase in the response times as the number of docker container instances with running VFs increase. The impact of the running VFs is visible so much that while it took the system a little above 8sec to instantiate 60 docker containers with running VFs, as in Fig. 6(b), it only took about half of that duration to instantiate the same amount of empty docker instances as shown in Fig. 6(a). These experiment results reveal that the higher the number of docker containers with running VFs are, the longer the time it will take to instantiate any new VF using the same system.

## VI. Concluding Remarks and Open Research Challenges

The ever-growing adoption of smart devices with enhanced capabilities is drastically modifying the industrial landscape. The required interconnectivity as well as the growing demands for low latency in data processing to enhance the current industrial networks (i.e., which have been originally designed for slowly/less-frequently changing environments) can not be overemphasized. In this context, the network slicing paradigm can be adopted and extended to accommodate the requirements of Industry 4.0. We have proposed a novel framework which aims at providing industrial network slices through federation over private/public infrastructures, supporting RAN slicing with multiple radio technologies, introducing multi-tenancy on industrial equipment, and endorsing the concept of control without ownership. The advantages of the proposed framework have been validated in three realistic use cases.

This paper aims at providing insightful directions towards the design and development of next-generation smart factories. Manifold open challenges are raised to effectively enable the envisioned system. Orchestration of network slicing is still under debate in the research communities. To boost its adoption over heterogeneous domains, great efforts in the standardization process should be carried out involving network operators, service providers, and industrial stakeholders. Furthermore, machine learning techniques can introduce notable advantages not only in the processing of industrial data, but also in the optimization and tuning of slicing management according to run-time operations. Novel hybrid approaches for industrial control systems should further investigate the interplay between local and global manufacturing optimizations. It is all the hope of the authors that this paper would stimulate further efforts in these directions among the concerned communities of academic and industrial researchers.

## Acknowledgment

This work was partially supported by the European Unions Horizon 2020 Research and Innovation Program through the MATILDA Project under Grant No. 761898. It was also supported in part by the 6Genesis project under Grant No. 318927 and by the Aalto 5G meets Industrial Internet (5G@II) project.

## References

[1] J. Q. Li, F. R. Yu, G. Deng, C. Luo, Z. Ming and Q. Yan, "Industrial Internet: A Survey on the Enabling Technologies, Applications, and Challenges," *IEEE Communications Surveys & Tutorials*, vol. 19, no. 3, third-quarter 2017, pp. 1504-1526.

[2] T. Taleb, "Toward carrier cloud: Potential, challenges, and solutions," *IEEE Wireless Communications*, vol. 21, no. 3, June 2014, pp. 80-91.

[3] A. Nakao, P. Du, Y. Kiriha, F. Granelli, A. A. Gebremariam, T. Taleb, and M. Bagaa, "End-to-end network slicing for 5g mobile networks," *Journal of Information Processing*, vol. 25, Feb. 2017, pp. 153-163.

[4] I. Afolabi, T. Taleb, K. Samdanis, A. Ksentini, and H. Flinck, "Network slicing and softwarization: A survey on principles, enabling technologies, and solutions," *IEEE Communications Surveys & Tutorials*, vol. 20, no. 3, third-quarter 2018, pp. 2429-2453.

[5] A. W. Colombo, S. Karnouskos, O. Kaynak, Y. Shi, and S. Yin, "Industrial cyberphysical systems: A backbone of the fourth industrial revolution," *IEEE Industrial Electronics Magazine*, vol. 11, no. 1, March 2017, pp. 6-16.

[6] X. Xu, "From cloud computing to cloud manufacturing," *Robotics and computer-integrated manufacturing*, vol. 28, no. 1, Feb. 2012, pp. 75-86.

[7] J. Wan, M. Yi, D. Li, C. Zhang, S. Wang, and K. Zhou, "Mobile services for customization manufacturing systems: An example of industry 4.0," *IEEE Access*, vol. 4, Nov. 2016, pp. 8977-8986.

[8] J. Wan, S. Tang, H. Yan, D. Li, S. Wang, and A. V. Vasilakos, "Cloud robotics: Current status and open issues," *IEEE Access*, vol. 4, June 2016, pp. 2797-2807.

[9] T. Taleb, K. Samdanis, B. Mada, H. Flinck, S. Dutta, and D. Sabella, "On multi-access edge computing: A survey of the emerging 5g network edge

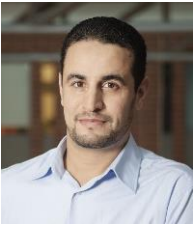

cloud architecture & orchestration," *IEEE Communications Surveys & Tutorials*, vol. 19, no. 3, third-quarter 2017, pp. 1657-1681.

[10] U. Raza, P. Kulkarni, and M. Sooriyabandara, "Low power wide area networks: An overview," *IEEE Communications Surveys & Tutorials*, vol. 19, no. 2, Second-quarter 2017, pp. 855-873.

[11] M. R. Palattella, M. Dohler, A. Grieco, G. Rizzo, J. Torsner, T. Engel, and L. Ladid, "Internet of things in the 5g era: Enablers, architecture, and business models," *IEEE Journal on Selected Areas in Communications*, vol. 34, no. 3, March 2016, pp. 510-527.

[12] D. Li, M. T. Zhou, P. Zeng, M. Yang, Y. Zhang, and H. Yu, "Green and reliable software-defined industrial networks," *IEEE Communications Magazine*, vol. 54, no. 10, October 2016, pp. 30-37.

[13] I. Afolabi, M. Bagaa, T. Taleb, and H. Flinck, "End-to-end network slicing enabled through network function virtualization," *Proc. 2017 IEEE Conference on Standards for Communications and Networking (CSCN)*. Helsinki, Finland, Sept 2017, pp. 30-35.

[14] T. Cruz, P. Simes, and E. Monteiro, "Virtualizing programmable logic controllers: Toward a convergent approach," *IEEE Embedded Systems Letters*, vol. 8, no. 4, Dec 2016, pp. 69-72.

[15] NFV-MAN, "Network functions virtualisation (nfv); management and orchestration," vol. 1.1.1, Dec. 2014.

Tarik Taleb Tarik Taleb received the B.E. degree (with distinction) in information engineering in 2001, and the M.Sc. and Ph.D. degrees in information sciences from Tohoku University, Sendai, Japan, in 2003, and 2005, respectively. He is currently a Professor with the School of Electrical Engineering, Aalto University, Espoo, Finland. He is the founder and the Director of the MOSA!C Lab. He is the Guest Editor-in-Chief for the IEEE JSAC series on network Softwarization and enablers.

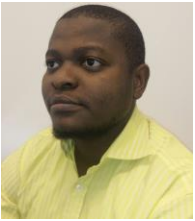

Ibrahim Afolabi Ibrahim Afolabi obtained his Bachelors degree from VAMK University of Applied Sciences, Vaasa, Finland, in 2013 and his Masters degree from the School of Electrical Engineering, Aalto University, Finland in 2017. He is presently pursuing his doctoral degree at the same university where he obtained his Masters degree from and his research areas include Network Slicing, MEC, network softwerization, NFV, SDN, and dynamic network resource allocation.

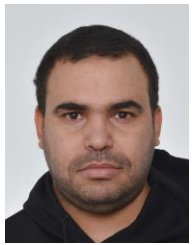

Miloud Bagaa Miloud Bagaa received the bachelors, masters, and Ph.D. degrees from the University of Science and Technology Houari Boumediene Algiers, Algeria, in 2005, 2008, and 2014, respectively. He is currently a Senior Researcher with the Communications and Networking Department, Aalto University. His research interests include wireless sensor networks, the Internet of Things, 5G wireless communication, security, and networking modeling.